# Spin direction controlled electronic band structure in two dimensional ferromagnetic CrI$_3$

*Peiheng Jiang[†,§], Lei Li[†,§], Zhaoliang Liao[‡,\*], Y. X. Zhao[⊥,#], and Zhicheng Zhong[†,\*]*

[†] Key Laboratory of Magnetic Materials and Devices & Zhejiang Province Key Laboratory of Magnetic Materials and Application Technology, Ningbo Institute of Materials Technology and Engineering, Chinese Academy of Sciences, Ningbo 315201, China

[‡]MESA+ Institute for Nanotechnology, University of Twente, PO Box 217, 7500 AE Enschede, The Netherlands

[⊥]National Laboratory of Solid State Microstructures and Department of Physics, Nanjing University, Nanjing 210093, China

[#]Collaborative Innovation Center of Advanced Microstructures, Nanjing University, Nanjing 210093, China

---

[\*] Email: z.liao@utwente.nl. Phone: +31 534892860.

[\*] Email: zhong@nimte.ac.cn. Phone: +86 574-86681852.

ABSTRACT: Manipulating physical properties using the spin degree of freedom constitutes a major part of modern condensed matter physics and is very important for spintronics devices. Using the newly discovered two dimensional van der Waals ferromagnetic $CrI_3$ as a prototypic material, we theoretically demonstrated a giant magneto band-structure (GMB) effect whereby a change of magnetization direction significantly modifies the electronic band structure. Our density functional theory calculations and model analysis reveal that rotating the magnetic moment of $CrI_3$ from out-of-plane to in-plane causes a direct-to-indirect bandgap transition, inducing a magnetic field controlled photoluminescence. Moreover, our results show a significant change of Fermi surface with different magnetization directions, giving rise to giant anisotropic magnetoresistance. Additionally, the spin reorientation is found to modify the topological states. Given that a variety of properties are determined by band structures, our predicted GMB effect in $CrI_3$ opens a new paradigm for spintronics applications.

Novel electronic phases in solid state materials can be driven by the spin degree of freedom due to the delicate interplay between spin, orbital, charge and lattice degrees of freedom.[1] To understand and control the spin is the key to many important electronic devices. In general, the spin ordering structures play a crucial role in determining the physical properties. For example, the metallic phase in correlated manganites is intimately related to long range ferromagnetic ordering, and a magnetic phase transition from ferromagnetism to para/antiferro-magnetism will trigger a metal-to-insulator transition.[2] The advanced multiferroicity results from the coupling between magnetic ordering and electric polarization.[3, 4] External magnetic field induced Zeeman spin splitting is usually used to induce magnetic ordering. However, the Zeeman spitting energy is extremely small with only an order of magnitude of $10^{-4}$ eV under a field of 1 T (see **Figure 1(a)**). Besides the Zeeman effect, magnetic field enforces a reorientation of the spin direction. Since the spin-orbit coupling (SOC) Hamiltonian term $\xi \boldsymbol{L} \cdot \boldsymbol{S}$ depends on the spin direction, electronic band structure may change when the spin direction is changed by an external magnetic field, said magneto band-structure (MB) effect. The MB effect can make spin direction a very useful knob to engineer the band structure and physical properties, in addition to the spin-order controlled properties. Although the importance of MB is very evident and efforts have been made to explore traditional ferromagnetic materials, it is frustrating that most of the ferromagnetic materials have too weak MB effect to produce interesting properties and functionalities. [5-9]

The big challenge arises from the fact that many solid state materials have either small spin-orbit coupling or high crystalline symmetry, leading to intrinsic weak MB effect. In order to make MB materials applicable for practical spintronics application, a significant change of band structure and thereby functionalities with different spin orientation is required. In this letter, we establish key criteria to realize giant MB effect (GMB) and demonstrate the GMB in a newly discovered two dimensional (2D) ferromagnet of monolayer CrI₃.[10-19] Our density functional theory calculations show that changing the spin direction of the monolayer CrI₃ remarkably modifies the band structures, including a direct-to-indirect bandgap transition, significant changes of Fermi surfaces and topological states. Our results reveal a new approach to manipulate properties by controlling the spin direction.

We use a 2D three-$p$-orbital ($p_x$, $p_y$, and $p_z$) toy model to illustrate the GMB effect. This model is described by Hamiltonian $H = H_0(k) + \left(\frac{\lambda}{2}\right)\sigma(\theta,\varphi) + \xi \boldsymbol{L}\cdot\boldsymbol{S}$.[20, 21] Here, $H_0(k)$ is the paramagnetic tight-binding Hamiltonian with matrix elements $H_{\alpha\beta} = \sum_{\boldsymbol{R}} t_{\alpha\beta}(\boldsymbol{R})e^{i\boldsymbol{K}\cdot\boldsymbol{R}}$. $t_{\alpha\beta}(\boldsymbol{R})$ represents a hoping integral from orbital $\alpha$ to orbital $\beta$ with lattice spacing $\boldsymbol{R}$, and $\boldsymbol{K}$ is the wave vector. The second term $\left(\frac{\lambda}{2}\right)\sigma(\theta,\varphi)$ describes an exchange splitting $\lambda$ with magnetization oriented along a specific direction $(\theta,\varphi)$ and $\sigma(\theta,\varphi)$ is the vector of Pauli matrices. The last term is SOC and the coefficient $\xi$ is the coupling strength. For a 2D ferromagnetic material, it is natural to define the out-of-plane direction as $z$-direction and in-plane as ($x$, $y$). The ferromagnetic exchange interaction upshifts spin down channel to higher unoccupied states. Meanwhile, the $|p_{z\uparrow}\rangle$ orbital (↑ means spin up) is split off by 2D confinement effect. Therefore, we mainly focus on $|p_{x\uparrow}\rangle$ and $|p_{y\uparrow}\rangle$ orbitals. We first consider a situation where the spin is oriented along out-of-plane direction (**M**//**c**). In this case, we have $\langle p_{x\uparrow}|\boldsymbol{L}\cdot\boldsymbol{S}|p_{y\uparrow}\rangle = -i$. As a result, the Hamiltonian has some nonzero off-diagonal terms and the bands at Γ become non-degenerate. A splitting energy of ~200 meV is found at Γ point when $\xi = 100$ meV is used for calculation (see Figure 1(b)). This energy splitting should be induced by SOC, because the splitting vanishes if the SOC is not included (e.g., $\xi = 0$). In contrast to the situation of **M**//**c**, the off-diagonal term $\langle p_{x\rightarrow}|\boldsymbol{L}\cdot\boldsymbol{S}|p_{y\rightarrow}\rangle$ becomes zero ($\rightarrow$ denotes spin up channel along in-plane direction) for **M**//***a***. Accordingly, the former splitting is removed and bands at Γ are degenerate as shown in Figure 1(c).

Switching a magnetic moment from easy axis (**M**//***c***) to hard axis (***M//a***) will cost magnetic anisotropic energy (MAE). The MAE is defined as $\Delta E = E_\rightarrow - E_\uparrow$, where $E_\rightarrow$ ($E_\uparrow$) is the total energy when the magnetic moment is oriented along in-plane (out-of-plane) direction. Integrating over all *k* points in the first Brillouin zone (BZ), the total energy is obtained, $E = \sum_i \int_{BZ} \varepsilon_i(k) f(\varepsilon_i - \varepsilon_F) dk$, where $f(\varepsilon_i - \varepsilon_F)$ is the Fermi-Dirac distribution. $\varepsilon_i$ and $\varepsilon_F$ are energy of *i*-th band and Fermi level, respectively.[21] The calculated MAE is on the order of $10^{-3}$ eV. This fact suggests that a giant change of bands by an order of magnitude of 0.2 eV can be achieved by overcoming only ~$10^{-3}$ eV MAE barrier. This GMB effect which arises from collective behavior in a 2D ferromagnet with strong SOC is much more significant than Zeeman effect in single free electron picture. The large crystalline anisotropy between in-plane and out-of-plane intrinsically exhibited by a 2D ferromagnet is very crucial for the GMB effect. To illustrate the role of symmetry, we extended our model to a non-layered 3D cubic system where the structure is more isotropic. It is found that the change of the band structure at Γ in 3D case is not observed because of equivalent spin orientation resulted from high symmetry.

As indicated by our toy model, we propose three criteria for discovering GMB materials: i) magnetically coupled electrons to obtain the collective behavior; ii) strong SOC to induce non-equivalent electronic behavior for different spin orientations; and iii) a reduced symmetry which has large crystalline anisotropy to maximize SOC anisotropy. These criteria suggest that the most promising candidate is a ferromagnetic material with strong SOC and high crystalline anisotropy.

An intuitive candidate is a 2D ferromagnetic material[10, 22] with strong SOC. Generally, long range ferromagnetic order mainly exists in 3$d$ transition metals and their compounds. However, 3$d$ elements have relative weak SOC and strong SOC can only be found in heavy elements, e.g., 5$p$ and 5$d$ elements. On the other hand, long-range magnetic order is prohibited in 2D materials due to the strong fluctuations at finite temperature according to Mermin-Wagner theorem.[23] Therefore, it is extremely challenging to discover a 2D ferromagnet with strong SOC.

Excitingly, the newly discovered 2D ferromagnet of monolayer CrI$_3$[10-19] fulfills all these criteria, providing us with an ideal candidate for realizing GMB effect. The CrI$_3$ possesses a graphene-like structure. The Cr atoms form honeycomb lattice and each Cr is surrounded by six I atoms (see Fig. 2).[21] The I$^{1-}$ ion has an orbital configuration of 5$s^2$5$p^6$, thereby having very strong SOC. Owing to octahedral symmetry, the five $d$-orbitals ($d_{xy}$, $d_{xz}$, $d_{yz}$, $d_{x^2-y^2}$, $d_{3z^2-r^2}$) of Cr are split into three $t_{2g}$ orbitals and two $e_g$ orbitals. Three $d$-electrons (3$d^3$) of Cr$^{3+}$ ion occupy the half-filled $t_{2g}$ orbitals and $e_g$ orbitals are empty.[24] These orbital configurations and 90º Cr-I-Cr bond lead to a ferromagnetic super-exchange interaction between two nearest neighboring Cr.[25-28] Together with existence of large magnetic anisotropy (MA) to suppress the spin fluctuation,[23] unique 2D ferromagnetic order can exist in monolayer CrI$_3$.[10, 28]. Since its discovery in 2017, many tantalizing magneto-transport and magneto-optic properties have already been realized in monolayer CrI$_3$.[13, 17-19]

There are several types of magnetic anisotropy,[29] e.g., shape anisotropy and magnetocrystalline anisotropy. The shape anisotropy favors an in-plane spin orientation for a 2D material. For CrI$_3$, the magnetocrystalline anisotropy favors an out-of-plane spin orientation.[16] Since the magnetocrystalline anisotropy is larger than shape anisotropy, the magnetocrystalline anisotropy dominates the MA behavior and results in an easy axis along out-of-plane direction.[10] The MAE of monolayer CrI$_3$ was calculated using first-principles density functional theory (DFT)[30, 31] (see calculation detail in supplementary materials[21]). Our results show that the easy axis is along out-of-plane and the MAE for spin oriented along in-plane *a* axis with respect to out-of-plane *c* axis is 0.98 meV/Cr (see Table I). Our obtained MAE is quantitatively consistent with the previously reported value of ~ 0.7 meV/Cr by Zhang *et al.*.[16] The little discrepancy arises from different calculation method: self-consistence in our calculation vs. non-self-consistence in their calculation. The spin direction of CrI$_3$ can be switched from out-of-plane to in-plane by applying an in-plane external magnetic field to overcome the MAE barrier. The question arises that whether the spin reorientation can significantly influence the electronic properties. Therefore, we calculated the electronic band structures with the magnetic moment along *c* (**M**//*c*) and *a* axis (**M**//*a*), respectively. The electronic band structures for these two different magnetic moment orientations are presented in Fig. 2(a) and (b), respectively.

One of the remarkable changes of electronic band structures is a direct-to-indirect bandgap transition when the magnetization is changed from **M**//*c* to **M**//*a*. In the case of **M**//*c*, the monolayer CrI$_3$ is a semiconductor with a direct gap of 0.91 eV, as shown in Figure 2(a). Since the standard DFT calculations will generally underestimate the bandgap, we further performed band structure calculation with Heyd-Scuseria-Ernzerhof (HSE) hybrid functional[32]. The bandgap from HSE hybrid functional is 1.64 eV.[21] Both the valence band maximum (VBM) and conduction band minimum (CBM) are located at Γ. The splitting energies between the two highest valence bands (VB) and the two lowest conduction bands (CB) at Γ are 174.50 meV and 64.42 meV, respectively. In addition, the band at Γ near VBM is mainly contributed by $p_x$ and $p_y$ orbitals of I atoms.[21] SOC is essential to lift the degeneracy of the band at Γ which otherwise becomes degenerate if the SOC is not included.[21] Such feature is consistent with the description in our toy model. When the magnetic moment is rotated from *c* axis to *a* axis, the energy splitting disappears for both VBM and CBM at Γ. Compared to the band structure for **M//c,** VB shifts down while CB shifts up at Γ. As a result, the CBM is shifted away from Γ to a region between M-K (see Fig. 2b) and the bandgap becomes indirect. Therefore, switching the magnetic moment from out-of-plane to in-plane causes a direct-to-indirect bandgap transition.

We propose that such a transition can be easily observed by measuring the photoluminescence. For a direct bandgap material, the electrons that are stimulated to CBM can directly relax to VBM without involving phonon interaction. Therefore, a strong photoluminescence is expected. On the other hand, the photoluminescence will be greatly suppressed for an indirect bandgap material. This significant difference in photoluminescence can be attested by measuring the photoluminescence under external magnetic field. Recently, Seyler *et al.* has reported circularly polarized photoluminescence in monolayer $CrI_3$ under linearly polarized excitation, indicating effective probe of magnetic order by photoluminescence.[13] In their experiments, only out-of-plane magnetic field was applied. It will be very interesting to further investigate the how the photoluminescence responses to in-plane magnetic field. Previously, photoluminescence measurement has been used to confirm a direct-to-indirect bandgap transition in 2D $MoS_2$ which was triggered by varying the thickness of $MoS_2$.[33, 34] The $CrI_3$ can exhibit some advantages over $MoS_2$ that the photoluminescence in $CrI_3$ can be reversibly and dynamically controlled by magnetic field.

Similar with conventional semiconductor such as Si, 2D semiconducting materials can also be doped with vacancies or substitutional impurities. Here, we employ rigid band shift to investigate the tuning of Fermi surface with spin reorientation in a doped monolayer $CrI_3$. Figure 3(a) and 3(b) show the Fermi surfaces of electron doped monolayer $CrI_3$ with different magnetization directions. The Fermi surface is found to change from one single ring around Γ point for **M**//***c*** (Figure 3(a)) to a shape of broad bean between M and K for **M**//***a*** (Figure 3(b)). When CrI$_3$ is hole doped, the Fermi surface is changed from one single ring (Figure 3(c), **M**//***c***) to the double rings (Figure 3(d), **M**//***a***) around Γ point. These changes are correlated to direct-to-indirect bandgap transition and different energy splitting behavior. Since Fermi surface plays a crucial role in determining the transport property of a material, the significant change of Fermi surface of doped monolayer CrI$_3$ will cause a strong spin direction dependent electronic transport. As a result, a giant anisotropic magnetoresistance (AMR) can be obtained by controlling direction of remnant magnetization with external magnetic field. Our Boltzmann transport theory with constant relaxation time calculations[35] demonstrate that the resistivity of CrI$_3$ monolayer for perpendicular magnetization ($\rho_\perp$, i.e., $\rho$(**M**//***c***)) is larger than that for in-plane magnetization ($\rho_\parallel$, i.e., $\rho$(**M**//***a***)). Here, the current used to measure the resistance is along in-plane ***a*** axis. This AMR effect which is due to the change of Fermi surface is fundamentally different from other kinds of AMR.[36] The AMR ratio is defined as $\Delta\rho/\rho_\perp = (\rho_\perp - \rho_\parallel)/\rho_\perp$. With a slight hole doping, an AMR ratio ~ 50% is obtained. A larger AMR ratio up to 70% can be achieved in heavily hole doped systems (see Figure S8 in suppmentary[21]). Our calculated GMB driven AMR in CrI$_3$ is significantly larger than conventional AMR in 3D ferromagnetic materials which is generally only a few percent,[8] making CrI$_3$ a promising candidate for spintronics application. In contrast to traditional GMR devices which are composed of alternated ferromagnetic and non-magnetic layers, the utilization of CrI$_3$ can potentially simplify the devices structures by avoiding multilayers stacking and constructing. It is worth to note that a similar AMR induced by band structure reconstruction has been experimentally realized in Sr$_2$IrO$_4$ recently.[37]

Given that the monolayer CrI$_3$ shares similar honeycomb structure with graphene and MoS$_2$, intriguing topological states are expected. Here, we further show that a topological phase transition can be induced by GMB effect in CrI$_3$. As indicated by blue circles in Figure 2(a), the second and third conduction bands cross each other without opening a gap along the K-Γ direction for **M**//***c***. By mapping out the band structure throughout the whole BZ with DFT-Wannier-TB method,[38-42] we show that six identical Dirac points exist in the BZ and each Dirac point is formed by a two-fold band crossing with linearly leading dispersion, namely, the coarse-grained fermions are massless Dirac fermions. Different from those in graphene with spin degeneracy, these Dirac points consist of only one spin channel. We further confirmed that all the Dirac points have nontrivial **Z**$_2$ topological charge by employing the formula $\gamma = \frac{1}{\pi}\oint_C \boldsymbol{A}(\boldsymbol{k}) \cdot d\boldsymbol{k}$ mod 2, where $\boldsymbol{A}(\boldsymbol{k}) = i\sum_\alpha \langle u_{\boldsymbol{k}}^\alpha | \nabla_{\boldsymbol{k}} | u_{\boldsymbol{k}}^\alpha \rangle$ is the Berry connection with $\alpha$ labeling the energy bands below the concerned Dirac point, and $C$ is any closed loop locally surrounding it. Due to the nontrivial character, we calculate the corresponding mid-gap edge states of our system along the zigzag direction as shown in Figure 4(a). The edge states all terminate at the projections of the Dirac points as indicated by the blue arrows. There are four projected Dirac points. The first and fourth ones are projected from single Dirac points, but the second and third ones are projected from two Dirac points (see detail in Fig. S14 in supplementary[21]). Accordingly, the second and third projected Dirac points are terminated by two edge states. When the magnetization is along ***a*-axis** (**M**//***a***), the former band crossing disappears and a bandgap is opened, as indicated by the blue circle in Figure 2(b). As the gapped phase is in the vicinity of a Dirac critical point, it is natural to calculate the Chern number by integrating the Berry curvature over the BZ.[21] The calculated Chern number is 1 when up to second conduction bands are considered, indicating that the system is modulated to a Chern topological insulator phase. The corresponding edge states along the zigzag direction are shown in Figure 4(b), where a chiral edge band connecting the second and third conduction bands is clearly found in consistence with the bulk-boundary correspondence that the Chern number dictates the number of chiral edge modes. It is worth to note that Hanke *et al.* recently reported a magnetization control of topological phase transition by spin–orbit torques and Dzyaloshinskii-Moriya interaction.[43]

In summary, we have proposed and demonstrated giant magneto band structure effect in a 2D ferromagnet of monolayer $CrI_3$. It is found that changing the spin direction of $CrI_3$ via an external field can significantly change electronic band structures, including a direct-to-indirect bandgap transition, Fermi surface and topological electronic states modification. Different resistance states can be obtained by switching the magnetization direction of $CrI_3$. Furthermore, the revealed direct-to-indirect bandgap transition can be used to develop magneto-optoelectronic device where the photoconductivity can be switched by an external magnetic field. Our results timely provide theoretical insight into underlying physics of monolayer $CrI_3$ and should motivate more experiment efforts. Since a variety of properties change with band structure, GMB materials can be used to develop new types of spintronics devices. In contrast to traditional spintronics devices which require stacking of several magnetic or non-magnetic materials, GMB materials will be able to simplify device structures and reduce the device size. Finally, we should emphasis that the proposed criteria to discover GMB materials can be applied straightforwardly into the other systems. Thanks to the fast development of materials epitaxial techniques, more 2D ferromagnetic materials can be artificially created in ultrathin films or heterostructural interfaces.[44] This provides lots of great opportunities for us to explore new GMB materials.

**Figures:**

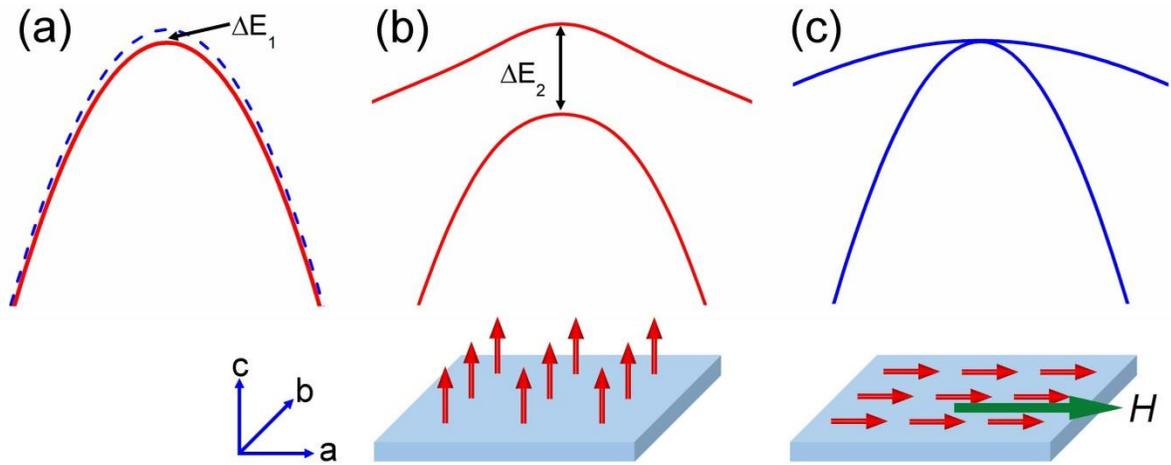

**Figure 1**. **Schematic view of Zeeman effect and giant magneto band structure effect. (a),** External magnetic field induced Zeeman splitting of a specific band. The energy splitting $\Delta E_1$ is in an order of magnitude of $10^{-4}$ eV under a 1 T magnetic field. **(b)-(c),** Calculated band splitting with toy model for a two dimensional system with **(b)** magnetization along out-of-plane $c$ axis (**M**//$c$), and **(c)** rearranged magnetization along in-plane $a$ axis (**M**//$a$) by applying magnetic field $H$. The energy splitting $\Delta E_2$ is in an order of magnitude of $10^{-1}$ eV, which is about $10^3$ times larger than Zeeman splitting.

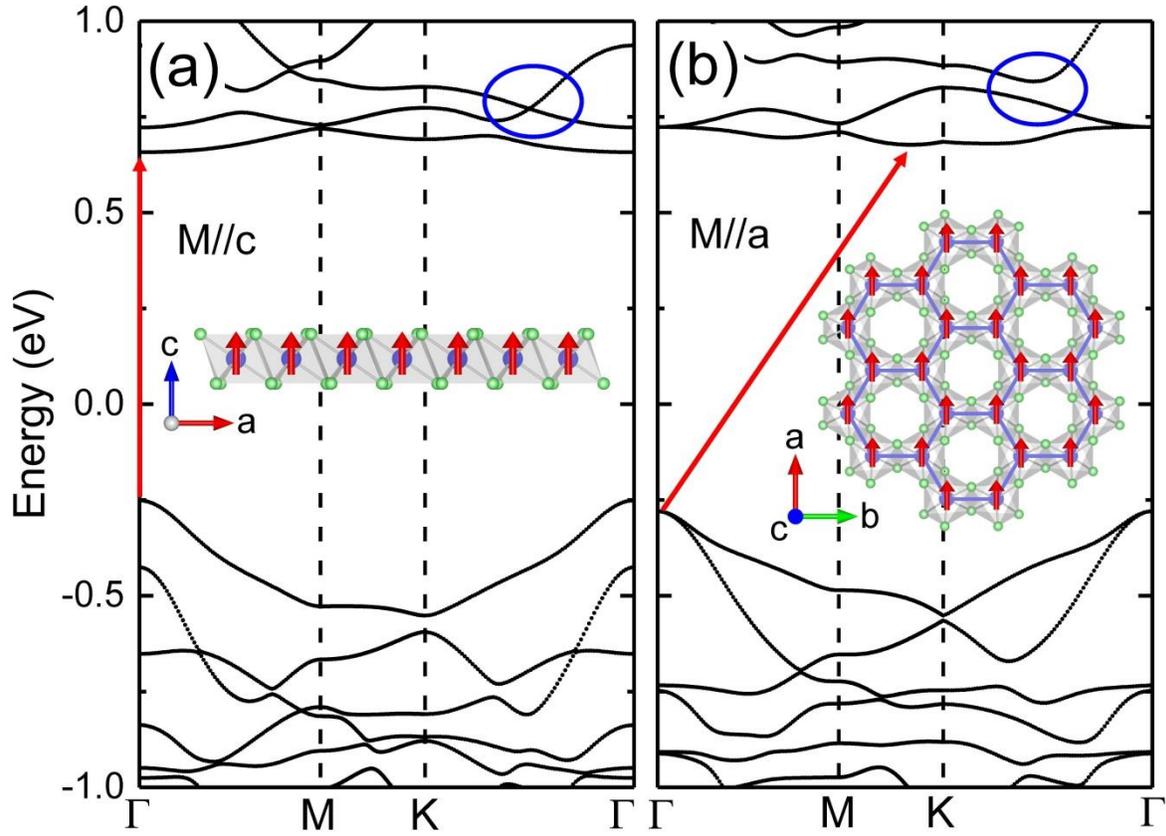

**Figure 2**. **The electronic band structures of monolayer CrI₃.** (**a**)**,** The band structure with magnetic moment along the out-of-plane *c* axis. (**b**)**,** The band structure with magnetic moment along the in-plane *a* axis. The red arrows depict the bandgap that is direct with **M**//*c* and indirect with **M**//*a*. The insets show the side view and top view of crystal structure with spin oriented along *c* axis and *a* axis, respectively. The blue and green balls represent the Cr and I atoms, respectively. The blue circles highlight the band-crossing when **M**//*c* which becomes gapped when **M**//*a*.

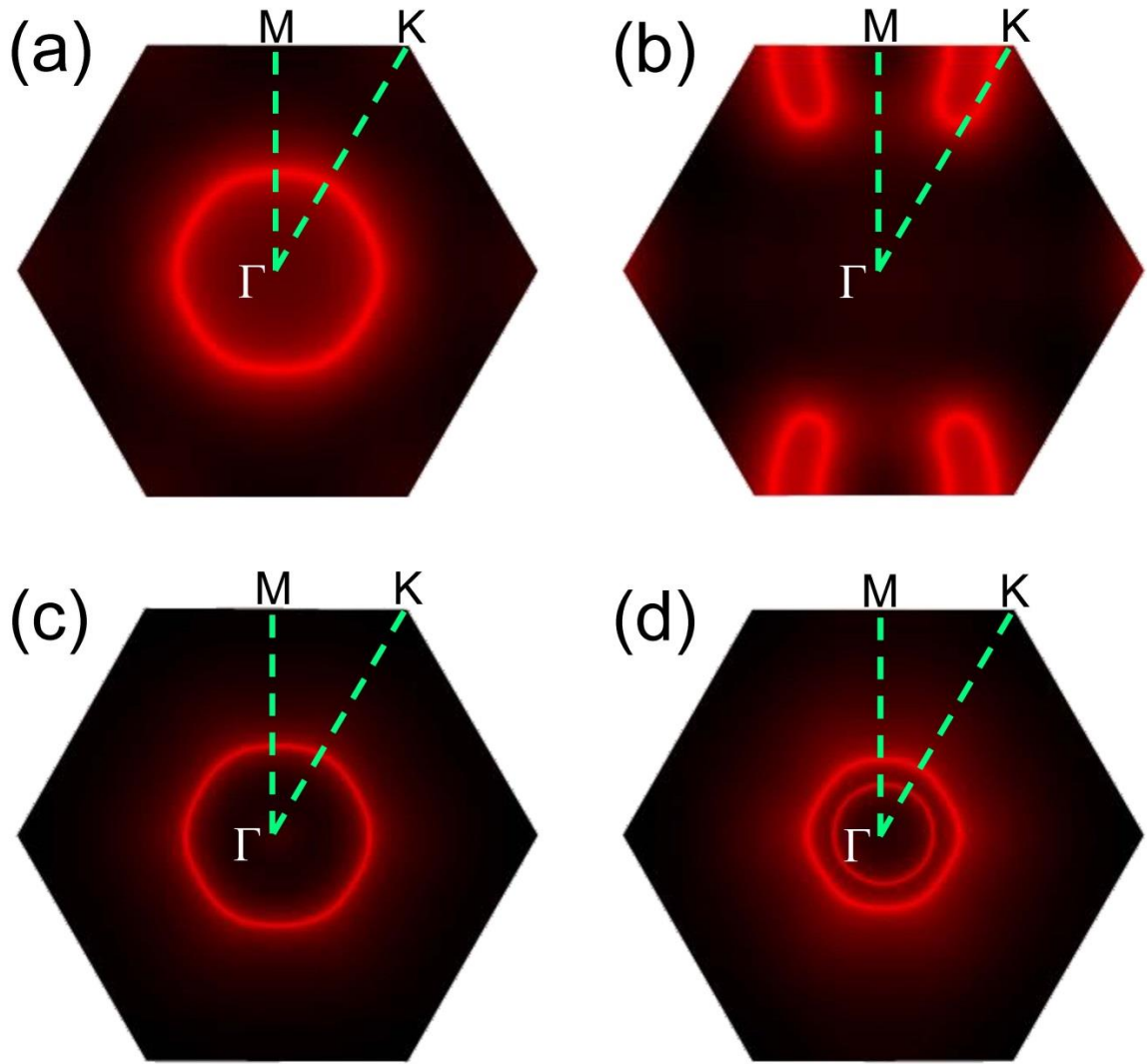

**Figure 3**. **Magnetization direction dependent Fermi surfaces of CrI$_3$ doped with 0.2 electron or hole**. (**a**), electron doping for **M**//***c***; (**b**), electron doping for **M**//***a***; (**c**), hole doping for **M**//***c***; (**d**), hole doping for **M**//***a***.

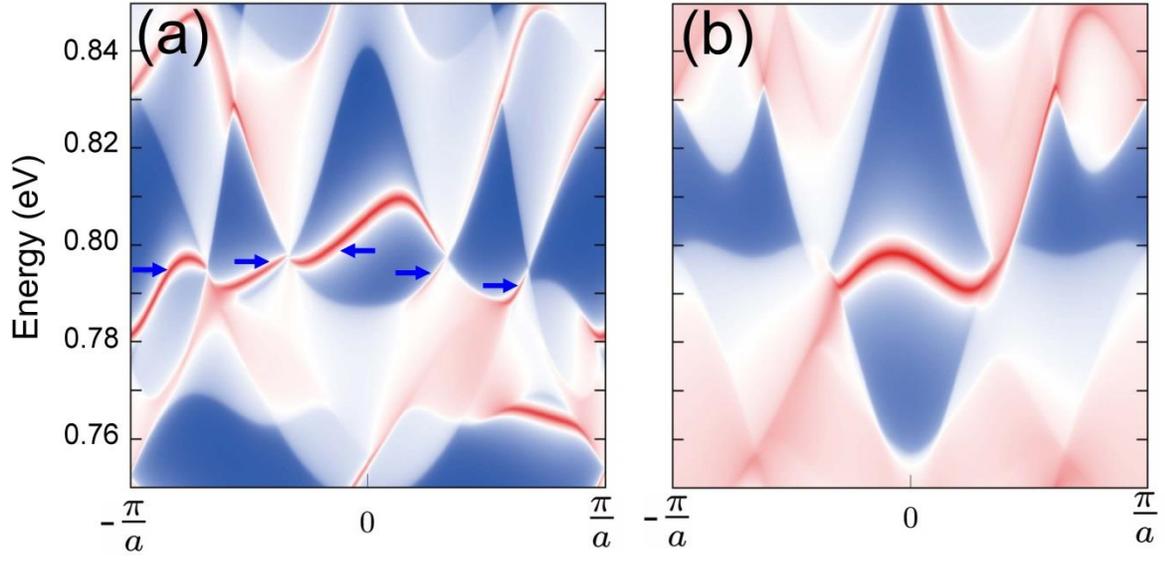

**Figure 4. The energy and momentum dependent local DOS (LDOS) on the edge of a semi-infinite sheet of CrI₃ along the zigzag direction.** (**a**), The LDOS for magnetic moment along the out-of-plane *c* axis. (**b**), The LDOS for magnetic moment along the in-plane *a* axis. The edge states are indicated by the blue arrows.

**Table 1. The MAE, bandgap, and energy splitting of monolayer CrI₃.** The calculations of MAE and energy splitting are performed with PBE functional, and bandgap calculations are performed with both PBE/HSE functionals.

|  | MAE (meV) | bandgap (eV) | | Γ split (meV) | |
| --- | --- | --- | --- | --- | --- |
|  |  | direct | indirect | VBM | CBM |
| M//*c* | 0.00 | 0.91/1.64 (Γ, Γ) | —— | 174.50 | 64.62 |
| M//*a* | 0.98 | —— | 0.96/1.69 (Γ, M-K) | 0.12 | 0.15 |

## ASSOCIATED CONTENT

**Supporting Information**. Details of the DFT calculated lattice structure, density of states (DOS), band structures, electrical conductivity, magnetic anisotropic energy and additional data.

## AUTHOR INFORMATION

**Corresponding Author**

*E-mail: zhong@nimte.ac.cn

*E-mail: z.liao@utwente.nl

**Present Addresses**

‡ Materials Science and Technology Division, Oak Ridge National Laboratory, Oak Ridge, Tennessee 37831, USA

**Author Contributions**

The manuscript was prepared through the contribution of all authors. All authors have given approval to the final version of the manuscript. Z.Z. supervised the project. P.J. and L.L. calculated and analyzed the results. §These authors contributed equally. All authors discussed the results and wrote the manuscript.

**Notes**

The authors declare no competing financial interests.

## ACKNOWLEDGMENT

P. J., L. L., and Z. Z. gratefully acknowledge financial support from the National Key R&D Program of China (2017YFA0303602), 3315 Program of Ningbo, and National Nature Science Foundation of China (11774360).

(42) Wu, Q.; Zhang, S.; Song, H.-F.; Troyer, M.; Soluyanov, A. A. *Comput. Phys. Commun.* **2018,** *224*, 405-416.

(43) Hanke, J. P.; Freimuth, F.; Niu, C.; Blugel, S.; Mokrousov, Y. *Nat. Commun.* **2017,** *8*, 1479.

(44) Geim, A. K.; Grigorieva, I. V. *Nature* **2013,** *499*, 419-425.
19

**Table of Contents**

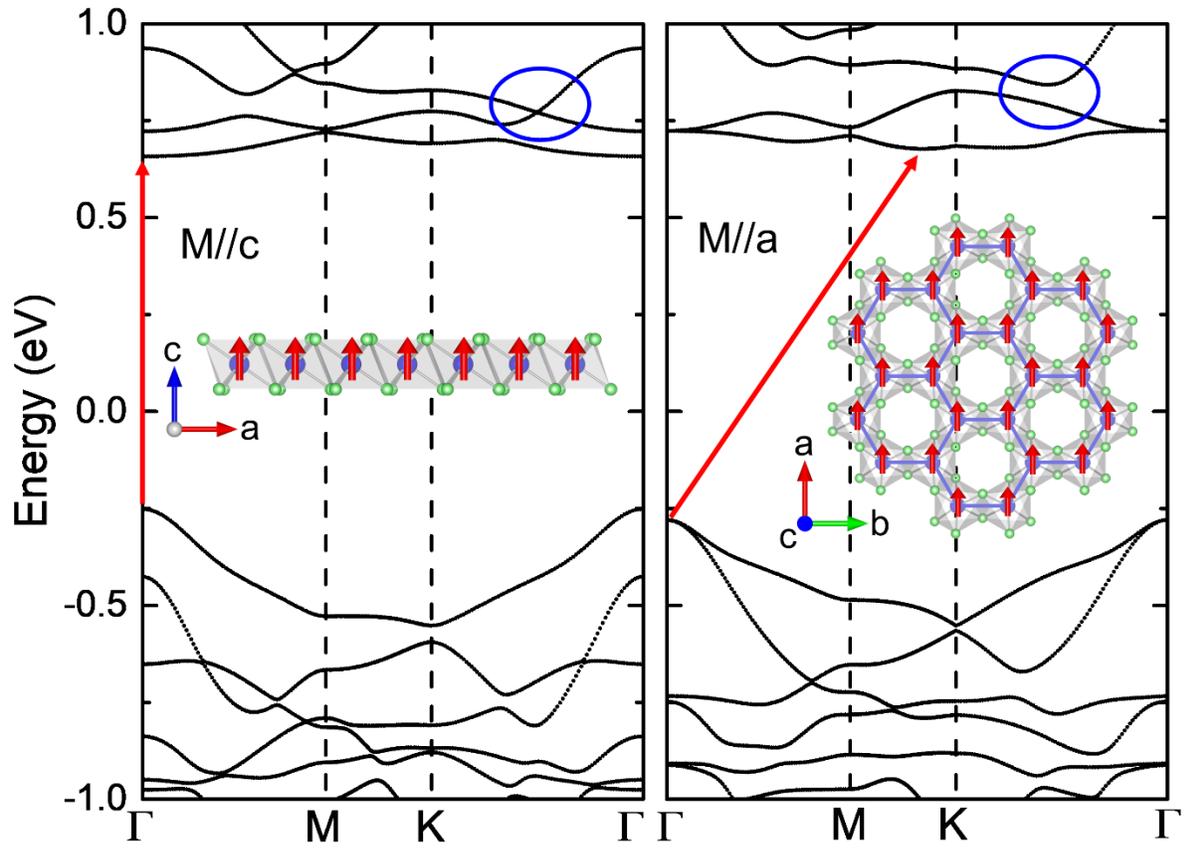